\documentclass[fleqn,10pt]{wlscirep}

\usepackage{subcaption}
\usepackage{float}

\title{A single-walker approach for studying quasi-ergodic systems}

\author[1,*]{Zilvinas~Rimas}
\affil[1]{Sidney Sussex College and Department of Chemistry,
	University of Cambridge, Cambridge, UK}
\affil[*]{zr219@cam.ac.uk}

\author[2]{Sergei~N.~Taraskin}
\affil[2]{St. Catharine's College and Department of Chemistry,
	University of Cambridge, Cambridge, UK}

\begin{abstract}
	The jump-walking  Monte-Carlo algorithm is revisited and updated to study the equilibrium properties of systems exhibiting quasi-ergodicity. 
It is designed for a single processing thread as opposed to currently predominant algorithms for large parallel processing systems.
	The updated algorithm is tested on the Ising model and applied to the lattice-gas model for sorption in aerogel at low temperatures, when dynamics of the system is critically slowed down. 
	It is demonstrated that the updated jump-walking simulations are able to produce equilibrium isotherms which are typically hidden by the hysteresis effect characteristic of the standard single-flip simulations.  
\end{abstract}

\begin{document} 
\flushbottom
\maketitle
\section*{Introduction} 
 \label{sec:Introduction}

The free-energy landscapes of complex interacting systems~\cite{Janke_2008} such as spin glasses~\cite{Binder_86,Marinari_1998:review,LandauBOOK}, Lennard-Jones clusters~\cite{Wales_2013}, supercooled liquids~\cite{Yamamoto2000}, proteins~\cite{Garcia2003}, lattice gas in porous media~\cite{Kierlik2001}, etc. are often characterised by many local minima which are separated by free-energy barriers. 
Understanding system's behaviour in such landscapes is a great challenge for experimental, analytical and numerical approaches~\cite{Chipot_Book2007}.  

Exponentially large number of states, the system can be in, restricts exact analysis to only small systems with the number of particles $N\sim 10^2$. 
The state space of the larger systems can be explored approximately by means of  numerical techniques such as Monte Carlo (MC) sampling. 
However, at sufficiently low temperatures the use of conventional MC methods is prohibited by exponentially long transition times over the free-energy barriers. 
At low temperatures, the system can be trapped in one of the local free-energy minima and thus it fails to ergodically explore the entire state space and achieve equilibrium.  
This behaviour is know as quasi-ergodic and it
has been, as well as a related broken-ergodicity problem, widely addressed in the literature  (see e.g. Refs.~\cite{Palmer_1982,Binder_86,Sharapov_2007,Chipot_Book2007,Neirotti_2000,Christ2010,Metzler_2014:review}).

Over last decades, numerous MC methods have been developed in order to alleviate the quasi-ergodicity problem~\cite{Iba2001,Bruce_2004,Christ2010,Iba_2014:review}. 
In this paper, we revisit one of the possible approaches developed in Refs.~\cite{Frantz1990,Frantz1992}, i.e. the jump-walking (JW) algorithm. 
As suggested by its name, the JW algorithm allows the system evolving in quasi-ergodic regime (e.g. at low temperature)  to jump into the states sampled from a different distribution (j-distribution) 
constructed for the system in ergodic regime (high-temperature) prior the beginning of the main simulation. 
This enables the algorithm to address the quasi-ergodicity problem, i.e. to overcome free-energy barriers by means of such jumps and ergodically explore the state space even at low temperatures. 
However, the original JW algorithm used for studying dynamics of small atomic  clusters~\cite{Tsai1991,Lopez1993,Lopez1996} had several drawbacks. 
First, it did not strictly satisfy the detailed balance~\cite{Geyer_1995,Hukushima1996REMconstr,Neirotti2000,Opps_2001} and thus did not guarantee to achieve correct (Boltzmann) limiting distribution for finite-length simulations.  
Second, the acceptance rate for jumps can be very low due to small overlap between high- and low-temperature distributions~\cite{Brown2003}. 
Third, high amount of slowly accessible computer memory on hard drive (rather than random-access memory (RAM) at that time) required for construction of the j-distribution  highly limited the applicability of the method only to relatively small systems~\cite{Tsai1991,Lopez1993,Lopez1996}. 
The main rout in further development of the JW algorithm and removing these constraints was through use of multiple walkers. 
In particular, the constraints of the JW method were addressed by developing several algorithms such as smart-walking~\cite{Zhou1997},  smart-darting~\cite{Andricioaei2001}, cool-walking~\cite{Brown2003},  replica-exchange method~\cite{Swendsen_1986,Hukushima1996REMconstr,Earl_2005} (REM) 
 (also known as parallel tempering~\cite{Marinari_1998:review}, multiple Markov chain method~\cite{Tesi1996}, parallel annealing~\cite{Kimura1991}) 
and combination of REM with other algorithms\cite{Sugita2000,Mitsutake2000}. 
The key feature of the majority of such approaches is in parallel running and exchange of  multiple non-interacting replicas of the original system~\cite{Marinari_1998:review,Neirotti2000,Iba_2014:review} in contrast to a single-replica simulation employed by the JW. 
Currently, the multiple-replica methods and their combinations with multicanonical algorithm dominate in studying quasi-ergodic behaviour. 

In this paper, we revisit the original JW algorithm, develop it further for the current state of computing technology and explore its new capabilities for studying quasi-ergodic behaviour of spin and lattice-gas models, which it has never been applied to. 
We demonstrate, that all the three major drawbacks of the original JW algorithm can be eliminated for a single-replica (single walker) version of the algorithm. 
This gives an opportunity to study low-temperature dynamics of lattice models of relatively large size ($\sim 10^5$ particles) using just a single-core simulations as contrasted to massively parallel ($\gtrsim 10^3$ cores~\cite{Vogel_2013,Vogel_2014}) simulations by REM or similar methods.

The structure of the paper is the following. 
The description of the original JW algorithm and its development  are presented in Sec.~\ref{sec:method}. 
The performance of the improved JW algorithm is discussed for the Ising  and lattice-gas models in Sec.~\ref{sec:Results}. 
The conclusions are given in Sec.~\ref{sec:Conclusions}. 
Some technical details are shown in Supplementary Information (SI).

\section*{Method}
\label{sec:method}
\subsection*{The original JW algorithm}
\label{sec:classicalJwalk}

Consider a system described by a Hamiltonian $H=H({Q})$ which is in equilibrium with the thermal bath at temperature $T$ (measured in energy units) where ${Q}$ denotes a state from the set $\{{Q}\}$ of all possible states the system can be in (canonical ensemble). 
Assume that the target temperature $T$ is low enough so that the system is quasi-ergodic at $T$. 
The JW algorithm~\cite{Frantz1990,Frantz1992} explores the state space by means of the standard MC sampling at $T$ augmented by jumps to the states which have been sampled at higher temperature $T'>T$. 
The temperature $T'$ is chosen high enough to ensure that the system is ergodic at $T'$. 
The states, to which the jumps are attempted to, are sampled according to the equilibrium Boltzmann distribution (thus constructing the j-distribution),
\begin{equation} 
P_{\text{B}}({Q},\beta')=\frac{e^{-\beta' H({Q})}}{\sum_{\{{Q}\}}e^{-\beta' H({Q})}}=
\frac{1}{{Z(\beta')}}e^{-\beta' H({Q})}~,
\label{eq:BoltzmanP} 
\end{equation} 
where ${Z(\beta')}$ stands for the partition function at the inverse temperature $\beta'=(T')^{-1}$. 
These additional jumps from states ${Q}_i$ (conventionally sampled at $T$) to states ${Q}_j$ (sampled at $T'$) allow for the ergodic sampling of the state space at temperature $T$ if they preserve the detailed balance, 
\begin{equation} 
\mathbb{P}({Q}_i\rightarrow{Q}_j)P_{\text{B}}({Q}_i,\beta)
=
\mathbb{P}({Q}_j\rightarrow{Q}_i)P_{\text{B}}({Q}_j,\beta)~,
\label{eq:detailedBalance} 
\end{equation} 
with $\mathbb{P}({Q}_i\rightarrow{Q}_j)$ being the transition probability between states ${Q}_i$ and ${Q}_j$. 

The transition probability, $\mathbb{P}({Q}_i\rightarrow{Q}_j)$, can be split into sampling and acceptance probabilities,
\begin{equation} 
\mathbb{P}({Q}_i\rightarrow{Q}_j)=
P_{\text{s}}({Q}_j,\beta')P_{\text{acc}}(\beta,\beta',{Q}_i,{Q}_j)~,
\label{eq:transProb} 
\end{equation} 
where $P_{\text{s}}({Q}_j,\beta')$ is proportional to the Boltzman probability at $T'$. The detailed balance condition given by Eq.~\eqref{eq:detailedBalance} imposes a constrain on the acceptance probability $P_{\text{acc}}$,
\begin{equation} 
\frac{P_{\text{acc}}(\beta,\beta',{Q}_i,{Q}_j)}
{P_{\text{acc}}(\beta,\beta',{Q}_j,{Q}_i)}=
e^{(\beta'-\beta)(H({Q}_j)-H({Q}_i))}~,
\label{eq:pAccConstraint} 
\end{equation} 
which is satisfied by the following choice~\cite{Frantz1990},
\begin{equation} 
P_{\text{acc}}(\beta,\beta',{Q}_i,{Q}_j)=
\text{min}(1,e^{(\beta'-\beta)(H({Q}_j)-H({Q}_i))})~.
\label{eq:metropolis} 
\end{equation} 
The standard Metropolis expression for $P_{\text{acc}}$ is recovered from Eq.~\eqref{eq:metropolis} as $T'$ goes to infinity ($\beta' \rightarrow 0$), since the j-distribution is then uniformly random. 
On the other hand, if the state space is sampled at the same temperature as that of the j-distribution ($\beta=\beta'$), the detailed balance is maintained with the acceptance probability independent of the state parameters, since the j-distribution itself would already encompass the required transition probability.

\subsection*{Development of the algorithm}
\label{sec:newJwalk}

Although the original JW algorithm successfully tackled the quasi-ergodicity problem, its implementation faced technical constraints at time of its creation. 
As mentioned in the Introduction, they were related to (i) inability to maintain the detailed balance, (ii) inefficiency of jumps between low- and high-temperature states with increasing temperature difference and (iii) low access rate of states in the j-distribution stored on hard drive.  
In this section, we discuss solutions to these problems and suggest an updated protocol for the JW algorithm, which addresses the mentioned issues.
Namely, we show that (i) the detailed balance can be maintained exactly within the new protocol presented below, (ii) the standard temperature-sequence  approach can be employed for the JW to increase the jump-acceptance probability and (iii) RAM can be used for accessing the j-distribution. 

We start with reviewing of the MC algorithms using ''jumpy'' protocols for dealing with quasi-ergodicity which can be considered as developments of the original JW algorithm aiming to improve its performance and  avoid its limitations and constrains.  
They can be broadly classified into two groups one based on multiple walkers (multiple replicas) and another one using one or two walkers.  
Multiple-replica methods with replica exchange are used to simulate a broad set of systems ranging  from spin glasses to bio-systems~\cite{Marinari_1998:review,Zuckerman_2011:review,Iba_2014:review,Ikebe_2016:review} while single-(two-)walker approaches are less popular and mainly used for modelling small atomic clusters~\cite{Tsai1991,Lopez1993,Lopez1996,Zhou1997,Andricioaei2001,Brown2003}. 
Multiple-walkers algorithms such as multiple-stage JW~\cite{Matro_1996},  REM~\cite{Swendsen_1986,Earl_2005} (parallel tempering~\cite{Geyer_1995,Hukushima1996REMconstr,Fiore_2010} which originated from simulated tempering~\cite{Marinari1992,Lyubartsev1992}) and annealed swapping~\cite{Opps_2001} run in parallel several walkers at temperatures covering the gap between high (ergodic) and low (target) temperature.  
This allows to increase the jump acceptance probability and preserve the detailed balance, although demanding massively parallel computations~\cite{Vogel2013,Vogel2014}.

Smart-walking (SW) method~\cite{Zhou1997}, in contrast to REM and similar to JW, uses only two  walkers running at high and low temperatures. In contrast to JW, SW before making a jump to a high-temperature distribution performs energy minimisation for high-temperature state and then makes a jump. 
SW significantly improves the acceptance probability for jumps and permits effective exploration of the energy landscapes especially for protein systems, however it does not maintain the detailed balance.
This drawback of SW (lack of detailed balance) is addressed by smart-darting (SD)~\cite{Andricioaei2001} algorithm which makes jumps only to neighbouring energy minima, although finding all relevant minima is problematic in complex landscapes. 
Cool walking~\cite{Brown2003} suggests a different method for achieving detailed balance in two-walker approach by means of statistical quench performed on configurations sampled by high-temperature walker. 

In the past the detailed-balance problem for the JW algorithm was addressed by two approaches. 
The first one~\cite{Matro_1996} requires multiple parallel simulations at high temperature $T'$, as well as one run at the target temperature, $T<T'$. 
The j-distribution is sampled at runtime by randomly choosing one of the systems simulated at $T'$ and attempting a transition to its current state. 
Use of several systems for sampling at high temperature prevents possible correlations, which may appear if sampling from the same simulation is done before it loses memory of the previously sampled state.
While this method does not use large amount of memory since the states are  sampled at runtime, it is based, similarly to REM, on multiple simulations running at the same time and it was replaced by well established REM maintaining the detailed balance~\cite{Neirotti2000}. 

The second approach~\cite{Frantz1990} is to perform the initial sampling of the j-distribution at $T'$ prior to the main simulation and save the recorded states. 
During the main simulation at $T$, a random state is periodically chosen from the saved j-distribution. 
The state could be accepted or rejected according to the Metropolis rule (Eq.~\eqref{eq:metropolis}). 
Due to insufficient amount of RAM, the set of states, $\{Q^{\text{rec}}(T')\}$, recorded at temperature $T'$,  was kept on hard-drive storage, thus severely limiting access speed and manipulation capabilities. 
In order to maintain the detailed balance, one had to ensure that the same state from $\{Q^{\text{rec}}(T')\}$ is not chosen twice~\cite{Frantz_1995,Neirotti2000}. 
Within this strategy, the probability of picking the same element in $\{Q^{\text{rec}}(T')\}$ more than once, decreases as the ratio of the initially recorded and eventually required states increases.  
However, even if significantly more states were recorded than the simulation at $T$ ultimately needed, this probability would remain finite and the detailed balance would be only approximately obeyed. 

Below we suggest a new algorithm which resolves the detailed-balance problem for a single-walker JW method. 
The key point of our approach is in using RAM for j-distribution which 
permits easy manipulations with recorded states. 
In particular, the recorded states stored in RAM can be easily removed once chosen for a possible jump-transition. 
Moreover, only necessary number of states from j-distribution for its unbiased sampling can be stored thus avoiding any storage overhead.  

A further challenge faced by JW as well as  REM and related methods arises due to the transition acceptance probability vanishing as the difference between temperatures $T$ and $T'$ increases~\cite{Neirotti2000}. 
This hinders access to the states sampled at $T'$ and thus prevents efficient equilibration of the system. 
The standard solution to this problem for the JW  algorithm~\cite{Frantz1995} is to run the simulation sequentially at several temperatures,  $\{T_\alpha\}~(\alpha =1,\ldots, n)$, with $T_1>T_2>\ldots >T_n$.  
The initial temperature $T_1$, is set high enough so that the system is fully ergodic at $T_1$ (see SI 1). 
Any two consecutive temperatures $T_\alpha$ and $T_{\alpha+1}$ are chosen in such a way that the Boltzmann distributions at these temperatures overlap, i.e. there exist states which the system can visit with non-vanishing probabilities at both temperatures. 
At each temperature $T_\alpha$, the target states ${Q}_j$ for the jump transitions are sampled from $\{Q^{\text{rec}}(T_{\alpha-1})\}$, and simultaneously the new j-distribution is constructed by recording the states to $\{Q^{\text{rec}}(T_{\alpha})\}$. 
This way the system remains ergodic at each successive temperature down to $T_n$. 
In case of  REM, a typical solution to the equivalent problem is to employ multiple different temperature replicas running concurrently~\cite{Swendsen1987,Hukushima1996}.

There has been a lot of work done to optimise the set of selected temperatures for REM~\cite{Katzgraber2006,Bittner2008} as well as the JW~\cite{Frantz1995,Frantz_1995} and other methods~\cite{Valentim_2014}. 
Typically, the temperatures are selected to follow geometric sequence~\cite{Kofke2002}. 
If the free-energy landscapes are sufficiently similar across the temperature range (which is not necessary the case e.g. for lattice polymers~\cite{Wust_2012} and Lennard-Jones clusters~\cite{Neirotti2000,Neirotti_2000}), the resultant overlaps between the subsets of the state space explored at different consecutive temperatures are comparable. 
This, in turn, leads to approximately the same acceptance probabilities for replica exchanges or JW jumps. 
However, in general, the free-energy landscape can be  significantly different for different problems (e.g. for proteins~\cite{Ikebe_2016:review} and  structural glasses~\cite{Cavagna_09:review,Ritort_03:review}). 
Therefore, the temperature-sequence optimisation strategy can depend on a particular system and preliminary simulations are often necessary to examine the subsets of available states at each temperature and optimise $\{T_\alpha\}$~\cite{Hukushima1996REMconstr}. 
The question of such optimisation is outside the scope of this work.
Instead, we provide a general-purpose multi-stage protocol for the JW  algorithm given an optimised set of temperatures $\{T_\alpha\}$.
Using this protocol the JW MC simulation run on a single thread can  ergodically explore the state space of the system exhibiting quasi-ergodic behaviour below a certain characteristic temperature $T_c$. 

The protocol is as follows:
\begin{itemize} 
	\item[(i)] Run the standard single-flip MC simulation at temperature $T_1>T_c$. 
Record the current state of the system at equal intervals between the regular MC steps, so that ${R}$ states are recorded into array $\{Q^{\text{rec}}(T_1)\}$ until the total required number, ${S}$, of MC steps  is executed.
	\item[(ii)] Decrease the temperature to $T_\alpha$, where $T_\alpha$ is the next element in the optimised set of temperatures $\{T_\alpha\}$, and run the standard single-flip MC simulation again.
	Choose ${R}$ step indexes at random from $1$ to  ${S}$ and once each of them is reached in the simulation, attempt a transition to a state randomly picked from $\{Q^{\text{rec}}(T_{\alpha-1})\}$ with the acceptance probability $P_{\text{acc}}$ defined by Eq.~\eqref{eq:metropolis}. 
Whether the transition is accepted or not, remove the suggested state from $\{Q^{\text{rec}}(T_{\alpha-1})\}$, so that it cannot be chosen again.
	In the meantime, continue to record the states as in (i) into the new array $\{Q^{\text{rec}}(T_{\alpha})\}$, which now contains an ergodic sample of the state space at $T_\alpha$.
	Proceed until ${S}$  MC steps are executed.
	\item[(iii)] Set $\{Q^{\text{rec}}(T_{\alpha})\}$ as the new $\{Q^{\text{rec}}(T_{\alpha-1})\}$ from which to sample the j-distribution. Repeat (ii) and (iii) until the final temperature $T_n$ in the list $\{T_\alpha\}$ is reached.
\end{itemize}

The algorithm provides us with two free parameters: ${R}$ and ${S}$. 
The constraint on precision naturally sets the required number of MC steps, ${S}$, to be completed.
The interval between the attempted jumps, ${S}/{R}$, is determined by the desired frequency of the jumps,  $\left({R}/{S}\right) \times P_{\text{acc}}$, and limited by how many states ${R}$ it is feasible to store in memory. 
For REM, it has been shown that given there were no computational constraints, equilibration accelerates as the replica-exchange frequency increases~\cite{Sindhikara2008,Sindhikara2010}. 
However, since the JW algorithm simulates only one copy of the system at any time, the jump transition, unlike the replica-exchange process in REM, requires to modify larger amounts of data associated with the state of the system. 
Computational cost of the transition is thus system dependent and could impose an upper bound on the optimal jump rate. 
Since the attempted states are removed from the storage independently of their acceptance, it is efficient to keep $P_{\text{acc}}$ as high as possible, minimising the number of stored states. 
Therefore, we are only free to adjust ${R}$ in order to set the required jump rate.

For the purpose of clarity, the above analysis assumes that the average acceptance probability $P_{\text{acc}}$ is the same for each $\{T_\alpha\}$. 
However,  in general, $P_{\text{acc}}$ depends on temperature and the constant jump-rate can be achieved  by adjusting the number of stored states $\{{R}_\alpha\}$ for each temperature $T_\alpha$.
All simulation parameters used for testing the algorithm are given in SI 1.

Finally, it is straightforward to generalise the JW method for  parameter-dependent Hamiltonians in a similar way to that used in the case of  multidimensional REM~\cite{Sugita2000mrem}. 
For a system described by a Hamiltonian $H=H({Q},\{\lambda_m\})$, where $\{\lambda_m\}$ is a set of parameters (e.g. chemical potential in the case of grand-canonical ensemble), an equivalent formalism leads to a generalised version of the expression for $P_{\text{acc}}$, 
\begin{equation} 
P_{\text{acc}}(\beta,\beta',{Q}_i,\{\lambda_m\},{Q}_j,\{\lambda'_m\})=
\text{min}\left(1,e^{\beta'(H({Q}_j,\{\lambda'_m\})-H({Q}_i,\{\lambda'_m\}))
-\beta(H({Q}_j,\{\lambda_m\})-H({Q}_i,\{\lambda_m\}))}\right)~.
\label{eq:genMetropolis} 
\end{equation} 
This allows us to sample the non-local transitions from a distribution constructed not only at a different temperature, but also based on Hamiltonian characterised by different values of parameters $\lambda'_m$.
An example of such analysis is presented in Sec.~\ref{sec:Aerogel}.

To summarise, the updated JW algorithm described above maintains the detailed balance by storing the j-distribution in RAM and, given an optimised set of temperatures, is able to explore ergodic behaviour of the system for relatively low temperatures at least for some lattice models as demonstrated in Sec.~\ref{sec:Results}. 

\section*{Results and discussion}
 \label{sec:Results}

In order to demonstrate the updated JW MC method and further clarify its performance, the algorithm has been implemented for two systems: a standard 3d Ising model and lattice-gas model for condensation of fluid in porous media. 
Both models, their implementation details and  simulation results are described in the following subsections.

\subsection*{Ising model}
 \label{sec:Ising}

The Ising model is one of the simplest models exhibiting phase transitions, making it well suited to test the performance of the JW algorithm.  
Since the memory required to record the states of such system grows linearly with the system size $N$, the total memory needed to store the multiple recorded states has been prohibitively high and the original JW method~\cite{Frantz1990,Frantz1992} has never been implemented for the Ising or other spin models.
Here, we demonstrate that the updated JW algorithm is capable of handling relatively large systems ($N \sim 10^{5}$) using only the memory of a regular contemporary personal computer and a single CPU core.

As a test example we study a 3d Ising model on a body centred cubic (bcc) lattice of $N=2 \times L^{3}$ ferromagnetically interacting spins (with periodic boundary conditions), where $L$ is the number of unit cells along the edge of a cubic sample. 
The dimensionality and lattice type are convenient for comparison with the lattice-gas models discussed in Sec.~\ref{sec:Aerogel}.

The Hamiltonian $\mathcal{H}_{\text{I}}$ of the Ising model with the interaction strength between two neighbouring spins $J>0$ ($J=1$) in external field $H$ (measured in units of $J$) is given by,
\begin{equation}
\mathcal{H}_{\text{I}} = -J \sum_{\langle ij \rangle} {\tau_i}{\tau_j}- H \sum^{N}_{i} {\tau_i}~,
\label{eq:isingHamiltonian}
\end{equation}
where the first sum is taken over all the nearest-neighbour spin pairs $\langle ij \rangle$ and $\tau_i \in \{+1,-1\}$ is the spin variable for a spin on site $i$. 

\begin{figure}[t]
	\centering
	\includegraphics[width=1\linewidth]{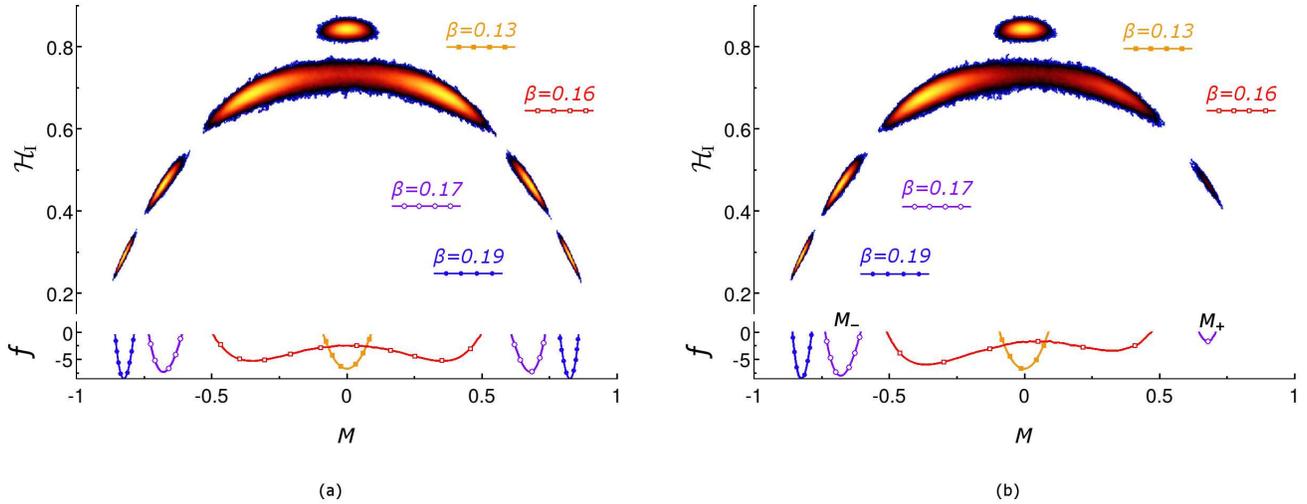}
	\caption{(a) Upper panel: the probability distribution of visits to macrostates characterised by energy (per spin) $\mathcal{H}_{\text{I}}$ (vertical axis) and magnetisation $M$ (horizontal axis) for the zero-field Ising model  described in the text. 
The distributions of visits at four labelled temperatures are shown. 
Two distributions for $\beta=0.17$ and $\beta=0.19$ consist of two disconnected regions each located at approximately the same horizontal level. 
The most visited macrostates are shown in yellow while the less visited ones are in the dark-blue. 
The white background refers to unvisited macrostates. 
Lower panel: Free-energy landscapes, i.e. $f(M)$, obtained for four distributions shown in the upper panel. The curves marked by different symbols refer to different temperatures labelled accordingly in the upper panel. 
(b) The same as in (a) but for non-zero external field $H=-5 \times 10^{-4}$. The values $M_+$ and $M_-$ in the lower panel indicate the positions of minima of $f(M)$ for $\beta=0.17$. The values of other parameters used in simulations are given in SI 1.
}
\label{fig:jwalk_ising}
\end{figure}

In order to visualise the multi-dimensional state space of the spin system, it is convenient to study its projection onto two dimensions representing magnetisation $M(\{\tau_i\})=N^{-1}\sum^{N}_{i} {\tau_i}$ (the order parameter) and energy $\mathcal{H}_{\text{I}}(\{\tau_i\})$ of a microstate $\{\tau_i\}$ (see the upper panels in Fig.~\ref{fig:jwalk_ising}).
In general, several spin configurations can have the same values of $M$ and $\mathcal{H}_{\text{I}}$ and it is convenient to define a macrostate ${Q}(M,\mathcal{H}_{\text{I}})$, with the degeneracy $g({Q}(M,\mathcal{H}_{\text{I}}))$, as a set of $g$ microstates with the same values of $M(\{\tau_i\})$ and $\mathcal{H}_{\text{I}}(\{\tau_i\})$. 
As MC sampling is running, the number of MC steps (or total time for kinetic MC~\cite{Bortz1975}) the system spends in a macrostate ${Q}(M,\mathcal{H}_{\text{I}})$ is recorded. 
In the ergodic regime, this quantity is proportional to the probability distribution for system to be in a certain macrostate and below we refer to it as distribution of visits. 
The probability to visit some  macrostates can be very small and domains of the state space containing such states (far tails of the probability distribution) cannot be explored by the JW algorithm for finite value of  parameter ${S}$~\cite{Neirotti_2000} (see Sec.~\ref{sec:newJwalk}).
The results for the distribution of visits are displayed on the $M$-$\mathcal{H}_{\text{I}}$ plane using the following colour-scheme. 
On the white background of unexplored state space, the most to the least visited macrostates are coloured in yellow to dark blue, respectively. 

Fig.~\ref{fig:jwalk_ising}a illustrates the results of the updated JW MC simulations on a $18 \times 18 \times 18$ bcc lattice ($N=11664$) for zero external field $H=0$ and several temperatures. 
At sufficiently high temperatures, all transitions between microstates are approximately equally likely, and therefore mainly the degeneracy of the macrostates determines which part of the state space is visited within the simulation time (see the single-peak distribution of visits around zero magnetisation for $\beta=0.13$ in the upper panel of Fig.~\ref{fig:jwalk_ising}a).
As temperature decreases and occupation of the low-energy states becomes exponentially more probable (according to Eq.~\eqref{eq:BoltzmanP}), the subspace of explored states shifts to the lower energies (cf. distributions at $\beta=0.13$ and $\beta=0.16$). 
Moreover, with decreasing temperature, the distribution of visits becomes bi-modal (see e.g. the distribution for $\beta=0.16$) reflecting the symmetry of the spin system with respect to the change of the spin orientation. 
This is due to the fact that the degeneracy of macrostates with approximately zero magnetisation decreases (with decreasing energy) more rapidly than of those with finite magnetisation. 
The two peaks in the distribution of visits become more pronounced with further decrease in temperature (cf. distributions at $\beta=0.17$ and  $\beta=0.19$) and at $T=0$  only two lowest energy states retain non-zero probability, each corresponding to all spins aligned either up or down.

This picture can be described by using the language of the free-energy  landscapes, referring to the dependence of the relative free energy $f(M)\equiv \beta (F(M)-F)$ solely on $M$, where
\begin{eqnarray}
F&=&-\beta^{-1} \ln Z = -\beta^{-1} \ln \left[ 
\sum_{\{\tau\}} e^{ -\beta \mathcal{H}_{\text{I}}}
\right]~, 
\label{eq:F} \\
F(M)&=&-\beta^{-1} \ln Z(M) =-\beta^{-1} \ln \left[  \sum_{\mathcal{H}_{\text{I}}}g({Q}(M,\mathcal{H}_{\text{I}}))e^{-\beta\mathcal{H}_{\text{I}}}
\right]~.
\label{eq:F_M}
\end{eqnarray}
The summation for free energy is taken over all microstates $\{\tau\}$ in Eq.~\eqref{eq:F} and over all macrostates $\left\{{Q}(M,\mathcal{H}_{\text{I}})\right\}$ with fixed magnetisation $M$ in Eq.~\eqref{eq:F_M}.  
It follows from Eqs.~\eqref{eq:F}-\eqref{eq:F_M} that the value of $f(M)=-\ln(P(M))$ is related to the probability $P(M)$ for the system to be in a state with fixed magnetisation, $P(M)=Z(M)/Z$. 
This probability can be estimated from the simulations by integrating the distribution of visits over the energies for fixed value of $M$.  
Thus obtained free-energy landscapes, $f(M)$, are shown in the bottom panels of Fig.~\ref{fig:jwalk_ising}. 

For temperatures above some critical value, $T_{\text{c}}$, the relative free energy $f(M)$ has a single minimum at zero magnetisation (see yellow line marked by solid squares in the bottom panel of Fig.~\ref{fig:jwalk_ising}a). 
At $T \approx T_{\text{c}}$, two symmetric minima at finite values of $M$ separated by a barrier at $M\simeq 0$ appear (see red line marked by open squares). 
With decreasing temperature,  the barrier between minima increases, the minima become deeper (see lines marked by the circles) and their positions tend to $M=\pm 1$ at zero temperature.

The picture described above cannot be seen by means of MC simulations  governed only by the standard single-flip mechanism, because the transitions through the free-energy barrier become exponentially rare with decreasing temperature. 
Eventually, within the standard single-flip MC simulations, the system settles down in one of the free-energy minima shown in Fig.~\ref{fig:jwalk_ising}a, thus violating the equilibrium and breaking ergodicity. 
This does not happen for the JW MC simulations for which the multiple-flip jump transitions to the states recorded at higher temperatures are possible.
These transitions enable exploration of all the states proportionally to the canonical distribution at the current temperature thus maintaining an  ergodic sampling of the state space and all free-energy minima present at sufficiently low temperatures become achievable within a single JW simulation.

In the presence of the finite external field, e.g.  $H<0$, (see Fig.~\ref{fig:jwalk_ising}b), similarly to the zero-field regime, the free-energy minima develop at positive, $M_+$, and negative, $M_-$, values of magnetisation (see e.g. the solid line marked by open circles for $\beta=0.16$ in the bottom panel of Fig.~\ref{fig:jwalk_ising}b). 
The negative external field breaks the symmetry and makes the minimum at $M_-$ deeper.
The JW algorithm samples the states at both free-energy minima according to the canonical probability distribution. 
This means that only the energy difference at these minima is significant for their relative occupation and the barrier between the minima is irrelevant for the JW sampling.   
As the macrostates around $M_+$ become ever less probable with decreasing temperature, the jump-transitions from this minimum to the minimum at $M_-$ are favoured over the reverse ones (see the asymmetric bi-modal distributions of visits at $\beta=0.16$ and $\beta=0.17$ in the upper panel of  Fig.~\ref{fig:jwalk_ising}b). 
Therefore, the amount of time that the system spends exploring the $M_+$ minimum gradually decreases with decreasing temperature (because the energy difference between minima increases) until it eventually vanishes completely for the precision set up within the JW algorithm (see the single-peaked distribution at $\beta=0.19$ in Fig.~\ref{fig:jwalk_ising}b).

\begin{figure}[t] 
	\centering
	\includegraphics[width=0.65\textwidth]{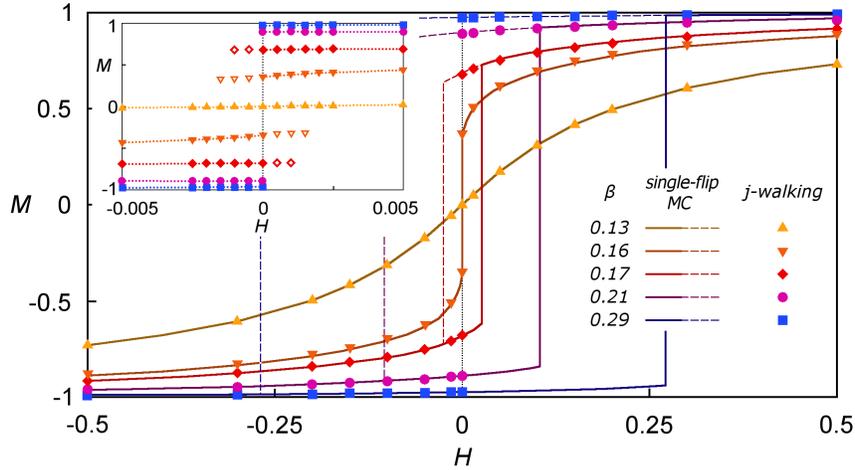} 
	\caption{Magnetisation reversal, i.e. the mean magnetic moment $M$ {\it vs} external field $H$, for the Ising model defined on bcc lattice with the same parameters as in Fig.~\ref{fig:jwalk_ising}, for several temperatures as marked in the legend. 
The data represented by solid (dashed) lines refer to increasing (decreasing) magnetic field and  were obtained by the single-flip conventional (kinetic)  MC simulations. 
For equilibrium isotherm at $\beta=0.13$ ($\blacktriangle$), the solid and dashed lines coincide. For $\beta=0.16$ ($\blacktriangledown$), a small hysteresis loop exists but is not distinguishable on the scale of the graph.   
The symbols represent results of the JW MC simulations.  
The inset shows magnified region around $H=0$. 
The solid (open) symbols in the inset refer to the mean magnetisation at energetically favourable (unfavourable) minimum of the free energy. 
The dotted horizontal lines in the inset serve as a guide for eye representing the value of magnetisation at the deepest minimum in the free-energy landscape. The dotted vertical line marks the transition at $H=0$.}
	\label{fig:2d_ising}
\end{figure}

An example of equilibrium behaviour of the Ising model obtained by means of the JW MC simulations contrasted with the results of the single-flip MC is presented in Fig.~\ref{fig:2d_ising}. 
The standard way to investigate $M(H)$ is to run a conventional single-flip MC simulations  at fixed temperature gradually incrementing the external field from sufficiently negative (so that $M\simeq -1$) to high positive values ($M\simeq 1$) and then back again. 
Under this protocol, a hysteresis loop is observed \cite{Sethna_review2006} (the areas between solid and dashed lines in Fig.~\ref{fig:2d_ising} for $\beta = 0.17, 0.19, 0.21$). 
In this hysteresis loop (in some systems found experimentally as well), the magnetisation of the sample  starts to align with the external field and then reverse in response to the change in $H$.
Such a hysteresis loop is still observed even if the change of the field takes place adiabatically slowly. 
This implies that the magnetic material is not in equilibrium
and is stuck in a state which it cannot leave on the single-flip MC (or experimental) time-scales with the
thermal energy available, i.e. the system gets stuck in a metastable state.
 
The JW MC algorithm provides a possibility to avoid trapping in metastable states, i.e. to remove the hysteresis loop, by sampling the state space according to the canonical equilibrium distribution. 
Indeed, Fig.~\ref{fig:2d_ising} (see the solid symbols corresponding to the JW data) shows a clear and expected phase transition at $H=0$ below the critical temperature $T_{\text{c}}$ (for bcc lattice, $T_{\text{c}}\simeq  0.157371(1)$~\cite{Lundow_2009:PhilMag}). 
Therefore, the JW MC simulations produce equilibrium isotherms, i.e.  $M(H)$, which are obfuscated by hysteresis phenomenon in the single-flip MC simulations or experiment. 
The inset in Fig.~\ref{fig:2d_ising} magnifies the transition region. 
In this region for fixed temperature (below critical) and $H$ close to zero, the JW MC simulations detect two minima of the free-energy, energetically favourable (solid symbols) and unfavourable (open symbols), which are also shown in lower panel of Fig.~\ref{fig:jwalk_ising}b. 
The probability to visit these minima follows the canonical equilibrium distribution and thus the accessibility of the energetically unfavourable minimum becomes exponentially small with increasing energy difference between the minima. 
Eventually, the energetically unfavourable minimum cannot be detected by the JW sampling with fixed precision. 
This occurs for relatively small deviations of the external field from zero, e.g. $|H|\sim 10^{-3}$ for $\beta= 0.17$. 
For the single-flip MC simulations, the height of the barrier between the free-energy minima matters and energetically unfavourable minima, in this case, act as traps, i.e. they become metastable states from which the system can escape only due to action of relatively large external fields, e.g. $H\sim 10^{-1}$ for $\beta = 0.17$.  

The behaviour of the Ising model described above is well established and served as a test for the JW MC simulations. 
As we can see from the results presented in Figs.~\ref{fig:jwalk_ising}-\ref{fig:2d_ising}, the  JW MC simulations reproduce all expected phenomena and, in addition,  allow the magnetisation equilibrium isotherms, i.e. $M(H)$, to be obtained for sufficiently low temperatures. 

\subsection*{Lattice-gas model}
 \label{sec:Aerogel}

In order to test the applicability of the updated JW MC algorithm to relatively large and complex systems, we have implemented the method for a lattice-gas model to study fluid sorption in porous media. 
Condensation of gas in porous materials occurs differently from that in free space, e.g. it takes place at a lower pressure (or chemical potential)  compared to the bulk saturation value~\cite{GreggBOOK1982,Gelb1999}.
While studies of fluids confined in single pores of simple geometry have clarified the mechanism for such shifted transitions~\cite{Evans1990,Gelb1999,Wallacher2004,Edison_2013,Handford_2013:pores}, the situation in real materials such as mesoporous glasses and silica gels, that consist of an interconnected network of pores of various shape and size, is still under active investigation~\cite{Wong1990,Detcheverry2003,Lambert2004,Detcheverry2005,Horikawa_11:review,Coasne_2013:review,Aubry2014,Handford_2014}. 

In this work, porous media is represented by two models: (i) a small lattice toy model for which exact numerical solution is available (see SI 2.2) and (ii) structural model of silica aerogel (see SI 2.3 for structural details of the model). 

In the lattice-gas model~\cite{Lee1952,Rothman1994,Kierlik1998,LandauBOOK}, each out of $N_{\text{tot}}$ lattice sites can be occupied by a fluid or a matrix particle, as described by the occupancy variables $\tau_i$ and $1-\eta_i$, respectively, which are equal to unity (zero) for occupied (unoccupied) sites. 
The matrix sites do not change their state, i.e. cannot be occupied by the fluid, and their concentration is quantified by porosity, $\phi={N_{\text{tot}}}^{-1}\sum^{{N_{\text{tot}}}}_{i} \eta_i$. 
In the simulation setup, the porous material is assumed to be an open system of $N=\phi {N_{\text{tot}}}$ pore sites connected to a reservoir of fluid particles. 
The number of the fluid particles $N_{\text{f}}= \sum_i^{{N_{\text{tot}}}} \eta_i\tau_i$ in the system and thus the energy $\mathcal{H}$ (see SI 2.1 for lattice-gas Hamiltonian) of the system can vary, but the chemical potential $\mu$, temperature $T$ and volume $V$ (or equivalently $N$) of the system are fixed. 
For a given set of $\mu$, $T$ and $N$, the system can be in a set of states which form a grand canonical ensemble described by the following grand partition function, 
\begin{equation} 
Z_{\mu}=\sum_{\{\tau\}} e^{ -\beta \tilde{\mathcal{H}} }~,
\label{eq:grand_part_func}
\end{equation}
where $\tilde{\mathcal{H}}={\mathcal{H}} -\mu N_{\text{f}} $, and the sum runs over all possible fluid occupation configurations $\{\tau\}$ of non-matrix sites, with the matrix variables $\eta_i$ being quenched. 
For the JW MC algorithm, the jump-transition acceptance probability (see Eq.~\ref{eq:genMetropolis}) for such system is given by the following expression,
\begin{equation} 
P_{\text{acc}}(\beta,\beta',{Q}_i,\mu,{Q}_j,\mu')=
\text{min}\left[1,e^{\beta'({\tilde{\mathcal{H}}}({Q}_j,\mu')-{\tilde{\mathcal{H}}}({Q}_i,\mu'))
-\beta({\tilde{\mathcal{H}}}({Q}_j,\mu)-{\tilde{\mathcal{H}}}({Q}_i,\mu))}\right]~,
\label{eq:aerogelMetropolis} 
\end{equation} 
where the non-local multiple-flip transitions from a current microstate of the system simulated at $\beta$ and $\mu$ to a state in the j-distribution previously sampled at $\beta'$ and $\mu'$ are allowed. However, to assure clarity and simplicity of the illustrative examples, the results presented below refer to a value of $\mu$ fixed within a single JW simulation.

The quantity of interest in our analysis is the mean equilibrium fluid density (order parameter), $\langle \rho \rangle =\langle N_{\text{f}}\rangle/N$ as a function of chemical potential and temperature. 
The angular brackets mean thermodynamic averaging over the states $\{\tau\}$ of the grand canonical ensemble, i.e.
\begin{equation}  
\langle \rho \rangle =  
Z_{\mu}^{-1} \sum_{\{\tau\}} \rho(\{\tau\}) e^{-\beta\tilde{\mathcal{H}}}~. 
\label{eq:mean_rho}
\end{equation}
The lattice-gas model is a discrete model and thus the density $\rho$ takes $N+1$ discrete values, $\rho(N_{\text{f}})=N_{\text{f}}/N=0, 1/N, \ldots,  (N-1)/N, 1$.  
Therefore, similarly to Eq.~\eqref{eq:F_M} it is convenient to rearrange the summation in Eq.~\eqref{eq:mean_rho} by first summing over all  macrostates  ${Q}(\rho,\tilde{\mathcal{H}})$ (constructed in the same way as ${Q}(M,\mathcal{H}_{\text{I}})$ in Sec.~\ref{sec:Ising}) with fixed number of fluid sites (i.e. fixed $\rho$) but variable energy $\tilde{\mathcal{H}}$ and then over fluid densities, $\rho$,
\begin{equation}  
\langle \rho \rangle = Z_{\mu}^{-1} \sum_{\rho} \rho \sum_{\{\tau(\rho)\}}  e^{-\beta\tilde{\mathcal{H}}(\rho)} = 
\sum_{\rho} \rho \sum_{\tilde{\mathcal{H}}}  P({Q}(\rho,\tilde{\mathcal{H}}))~. 
\label{eq:mean_rho2}
\end{equation}
The probability, $P({Q}(\rho,\tilde{\mathcal{H}}))$, introduced in Eq.~\eqref{eq:mean_rho2}, to find the system in a certain macrostate ${Q}(\rho,\tilde{\mathcal{H}})$ is given by 
\begin{equation}  
P({Q}(\rho,\tilde{\mathcal{H}})) = Z_{\mu}^{-1} g({Q}(\rho,\tilde{\mathcal{H}}))  e^{-\beta\tilde{\mathcal{H}}(\rho)}~, 
\label{eq:pdf_q}
\end{equation}
where $g(\rho,\tilde{\mathcal{H}})$ stands for the degeneracy of macrostate ${Q}(\rho,\tilde{\mathcal{H}})$, 
and thus the probability of the system to be in a state with a particular $\rho$ is $P(\rho)=\sum_{\tilde{\mathcal{H}}}  P({Q}(\rho,\tilde{\mathcal{H}}))$.

\begin{figure}[t]
	\centering
	\includegraphics[width=1\linewidth]{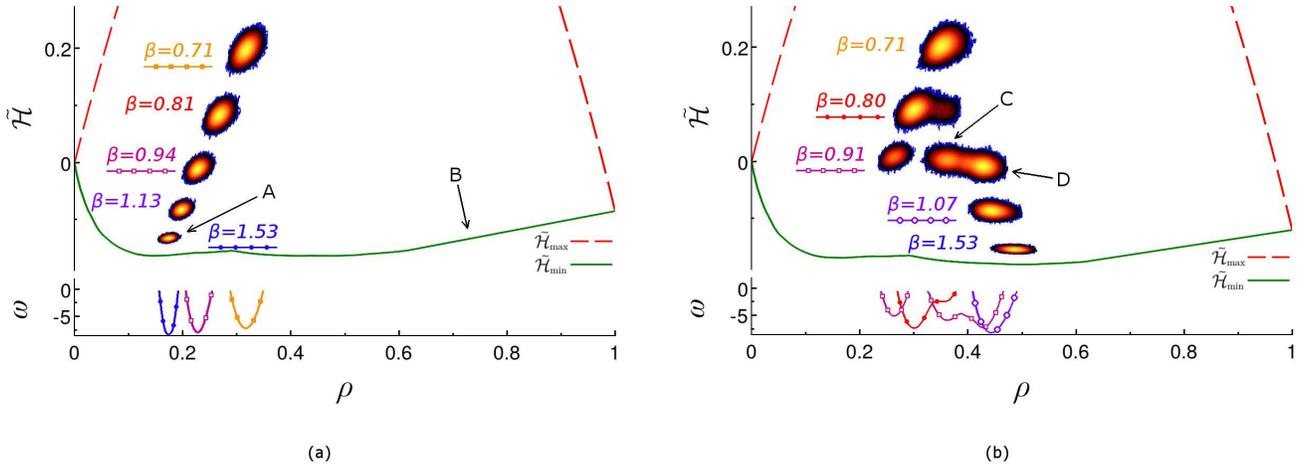}
	\caption{Upper panel: The probability distribution of visits of macrostates characterised by relative energy per pore site, $\tilde{\mathcal{H}}$ (vertical axis),  and density $\rho$ (horizontal axis) for the lattice-gas model applied to aerogel sample as described in the text. 
The distributions at five labelled temperatures for (a) $\mu=-4.18$ and (b) $\mu=-4.145$ are shown using the same colour scheme for density of visits as in Fig.~\ref{fig:jwalk_ising}. 
The estimates for the upper and lower boundaries of the state space are shown by the dashed (red) and solid (green) lines. 
The arrows A, B, C and D point to the states characterised by different fluid densities (see SI 2.4 for graphical illustrations of fluid configurations).   
Bottom panels: Grand-potential landscapes, i.e. $\omega(\rho)$, obtained for  
three temperatures: (a) $\beta=0.71$ ({\tiny $\blacksquare$}), $\beta=0.94$ ({\tiny $\square$}) and $\beta=1.53$ ($\bullet$); 
(b) $\beta=0.80$ ($\bullet$), $\beta=0.91$ ({\tiny $\square$}) and $\beta=1.07$ ($\circ$) (see SI 1 for the values of other parameters). 
}
\label{fig:jwalk_aerogel}
\end{figure}

Similarly to $F(M)$ in the case of the Ising model (see Sec.~\ref{sec:Ising}), the grand-potential landscape can be defined as the $\rho$-dependence of the grand potential $\Omega(\rho)=-\beta^{-1} \ln Z_{\mu}(\rho)$, where $Z_{\mu}(\rho) =Z_{\mu} P(\rho)$.
The shape of $\Omega(\rho)$ can describe possible phases of the system. 
For example, if $\Omega(\rho)$ has one minimum then the system is characterised by a single phase. 
However, if two minima appear in $\Omega(\rho)$ this might be an indication of coexistence of two phases with different densities. 
In order to visualize the grand-potential landscape, similarly to $f(M)$ in Sec.~\ref{sec:Ising}, we plot $\omega(\rho)\equiv \beta (\Omega(\rho)-\Omega)=-\ln(P(\rho))$~\cite{Woo2003} (calling this quantity as grand-potential landscape), where $\Omega=-\beta^{-1} \ln Z_{\mu}$ is the total grand-potential of the system (see the bottom panels in Fig.~\ref{fig:jwalk_aerogel}).

The state space of the lattice-gas model can be sampled by means of the JW MC algorithm in a similar way as that of the Ising model. 
The distribution of visits of different macrostates is displayed in the  $\rho$-$\tilde{\mathcal{H}}$ plane (see the upper panels in  Fig.~\ref{fig:jwalk_aerogel}), where  the same colour-scheme is used as in Fig.~\ref{fig:jwalk_ising}. 
Approximate high- and low-temperature boundaries of the state space are also computed (see SI 3 for details) and displayed for reference in the upper panels of Fig.~\ref{fig:jwalk_aerogel} (the red (dashed) and green (solid) lines, respectively). 

Fig.~\ref{fig:jwalk_aerogel} presents the results of the JW MC simulations for sorption of fluid in a model sample of aerogel for two different values of $\mu$ and several temperatures. 
At low temperatures for both values of $\mu$ shown in Fig.~\ref{fig:jwalk_aerogel}, the grand-potential landscape has two minima, as can be inferred from the shape of the low-temperature boundary, $\tilde{\mathcal{H}}_{\text{min}}(\rho)$ (the solid green line in the upper panels). 
The low-density minimum (e.g. at $\rho\simeq 0.15$ for $\mu=-4.18$) represents the fluid distribution in the system for which the sites occupied by fluid are concentrated only around and nearby the quenched matrix sites (see SI 2.4). 
The minimum at the higher values of $\rho$ corresponds to the  aerogel sample  filled with fluid, and only the added surface layers left unoccupied (see SI 2.4). 
As temperature is decreased, the explored part of the state space shifts down until the system eventually settles in a minimum (see the sequences of distribution of visits corresponding to gradually decreasing temperature in the top panels of Fig.~\ref{fig:jwalk_aerogel}). 
For sufficiently low or high values of $\mu$ the system straightforwardly descents to the low- or high-density minimum, respectively (the descent to the low-density minimum is shown in Fig.~\ref{fig:jwalk_aerogel}a). 
However, at the intermediate values of $\mu$  (see Fig.~\ref{fig:jwalk_aerogel}b) the state space sampled by the JW MC simulation splits into several regions and thus the distribution of visits has several peaks (see e.g. the distribution at $\beta=0.91$ in the top panel of  Fig.~\ref{fig:jwalk_aerogel}b).
In this case, the right peak in the distribution (see colour-map for $\beta=0.80$ in  Fig.~\ref{fig:jwalk_aerogel}b) becomes dominant with decreasing temperature and eventually the system settles down in the high-density minimum of the grand-potential.
The single-flip MC simulations would not be able to follow such equilibration of the system for $\mu=-4.145$. 
This is due to development of the large entropic barrier between two minima leading to significant slowing down of single-flip dynamics. 
Without the jump-transitions provided by the JW MC simulations, the region of the state space corresponding to the high-density peak could not be sufficiently visited and system would remain in the low-density state, i.e. in the metastable state.

Fig.~\ref{fig:jwalk_aerogel}b also shows the existence of the secondary split of the high-density peak in the distribution of visits into two peaks (see the two neighbouring bright spots on the right in the colour map for $\beta=0.91$) corresponding to two minima in the grand-potential landscape at $\rho\simeq 0.35$ and  $\rho\simeq 0.45$ (see the solid line marked by the open squares in the bottom panel). 
This feature (two high-density peaks in the distribution of visits) is an aerogel-sample specific (see SI 2.4). 
In addition to these two high-density configurations at $\beta=0.91$, there is a low-density one with $\rho \simeq 0.25$ (see the corresponding minimum of $\omega(\rho\simeq 0.25)$ in the bottom panel). 
For this configuration, mainly the pore sites surrounding the matrix sites are occupied  by fluid.

\begin{figure}[t] 
	\centering
	\includegraphics[width=1\linewidth]{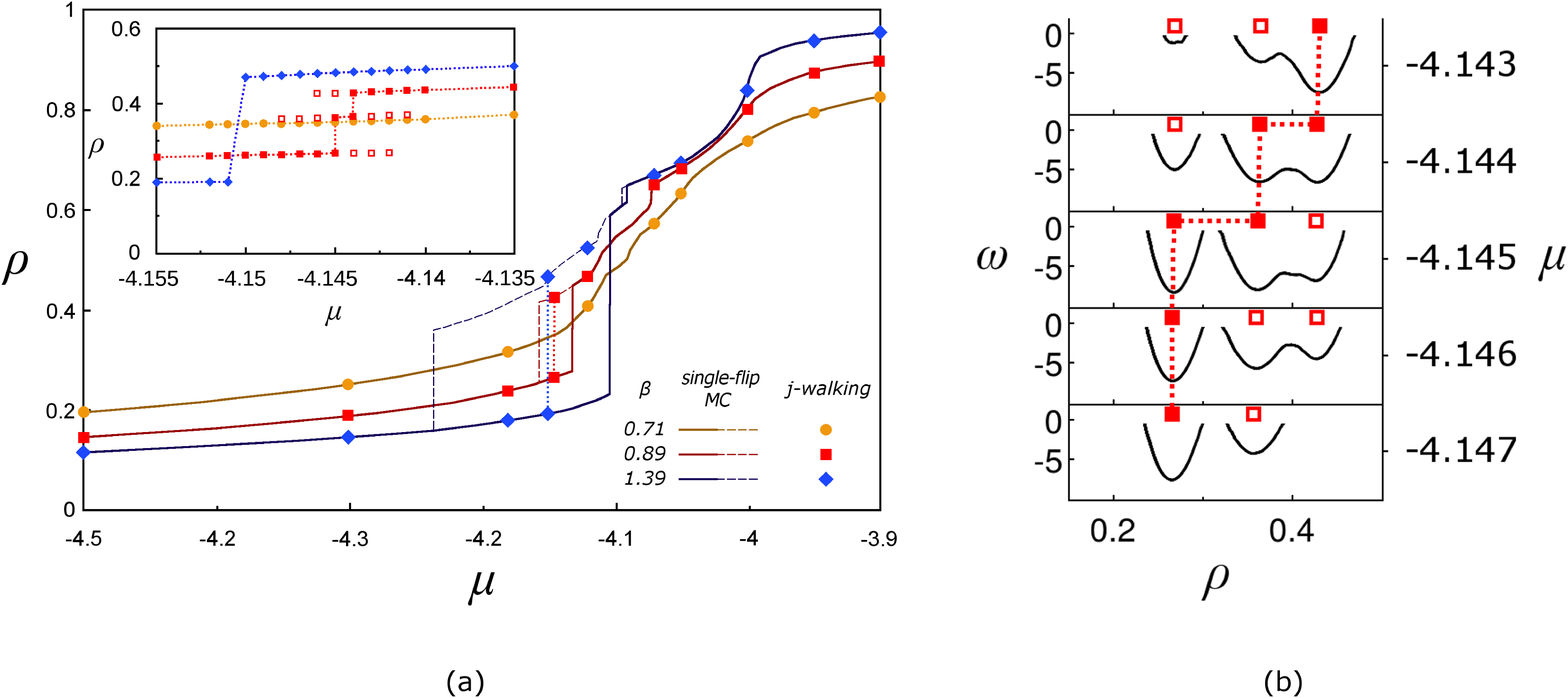}
	\caption{
(a) The dependence of the mean fluid density $\rho$ in structural model of aerogel {\it versus} chemical potential $\mu$ for several temperatures as marked in the figure.  
The results of the single-flip conventional MC simulations for adsorption (increasing $\mu$) and desorption (decreasing $\mu$) are shown by solid and dashed lines, respectively. 
The symbols refer to results of the JW MC simulations. 
For equilibrium isotherm at $\beta=0.71$,   the solid and dashed lines coincide with each other and with the solid circles.  
The inset shows magnification of the isotherms obtained by the JW MC simulations for the values of $\mu$ where the discontinuous transitions occur at low temperatures. The solid (open) symbols correspond to the deepest minimum  (other minima) in the grand-potential landscape, $\omega(\rho)$. 
(b) Grand potential landscapes, $\omega(\rho)$, for $\beta = 0.89$ and several values of $\mu$ specified on the right from each panel. 
The solid and open squares mark the deepest  and other minima of $\omega(\rho)$, respectively, in the same way as in (a). 
}
	\label{fig:2d_aerogel}
\end{figure}

The JW MC simulations performed for fixed temperature and different values of chemical potential  provide equilibrium isotherms, i.e. $\rho(\mu)$,  for sorption in aerogel (see the data points labelled by circles, squares and diamonds in Fig.~\ref{fig:2d_aerogel}a).  
The solid symbols refer to the densities corresponding to the deepest minimum of $\omega(\rho)$ while the open symbols (see the inset in Fig.~\ref{fig:2d_aerogel}a) are related to other minima of $\omega(\rho)$ available for given values of $\mu$ and $\beta$. 
A typical evolution of the grand-potential landscape with chemical potential for  relatively low  temperature ($\beta = 0.89$) is shown in  Fig.~\ref{fig:2d_aerogel}b. 
In two bottom panels ($\mu=-4.147$ and $-4.146$), the low-density minimum (marked by solid square) is the deepest one and system mainly has the density corresponding to this minimum. 
One (for $\mu=-4.147$) and two (for $\mu=-4.146$) higher-density minima marked by open squares are weakly occupied in equilibrium. 
They are separated by a large barrier (not fully shown) from the low-density minimum and cannot be detected by the single-flip MC simulations due to large height of this barrier as compared to temperature. 
For $\mu = -4.145$, the low- and intermediate-density minima become of approximately the same depth (they both marked by solid squares) and the system, in equilibrium, can be in one of them with approximately equal probability but, again, this cannot be detected by the single-flip MC simulations due to high barrier between the minima. 
In the relatively narrow range of chemical potential, $-4.145 \leqslant \mu \leqslant -4.144$, the most likely fluid density is related to the middle minimum of $\omega(\rho)$ (see the inset in Fig.~\ref{fig:2d_aerogel}a)  which becomes approximately of the same depth as the high-density minimum at $\mu \simeq -4.144$ (both minima are marked by solid squares in Fig.~\ref{fig:2d_aerogel}b). 
For greater values of chemical potential, $\mu \geqslant -4.144$, the system in thermal equilibrium is in the high-density state (see the top panel in Fig.~\ref{fig:2d_aerogel}b where the high-density minimum is marked by solid square and two other states with lower density by open squares).  

In contrast to the JW MC simulations, both experimental studies~\cite{Aubry_2013,Aubry2014} and single-flip MC simulations~\cite{Sarkisov_2000:Langmuir,Sarkisov_2001:PRE} are not able to access the equilibrium for sufficiently low temperatures due to presence of the hysteresis effect.  
Indeed, the results of the single-flip MC simulations for adsorption (solid lines in Fig.~\ref{fig:2d_aerogel}a) and desorption (dashed lines) for the same aerogel model exhibit  hysteresis (area between solid and dashed lines) for $\beta=0.89$ and $\beta=1.39$, although both adsorption and desorption curves coincide producing the equilibrium isotherm (solid line marked by circles representing the JW MC data) for $\beta=0.71$. 
Therefore, obtaining the equilibrium isotherms for sorption in aerogel and other porous materials is often problematic, and requires techniques such as mean-field scanning curves~\cite{Detcheverry2003}. 
However, despite those efforts conclusive answers have not yet been obtained to even the key questions such as about the existence of the discontinuous phase transition in aerogel samples of different porosity~\cite{Aubry2014}.
The JW algorithm is designed to avoid hysteresis and thus can be an invaluable tool in detecting phase transitions and resolving such problems. 
A detailed analysis of equilibrium sorption isotherms in aerogel samples of different porosity is outside the scope of this paper and will be presented elsewhere.

\section*{Conclusions}
 \label{sec:Conclusions}

To conclude, we have revisited the JW algorithm~\cite{Frantz1990} which has been originally designed to tackle quasi-ergodicity problem, a known problem in dynamics of interacting particles.
However, due to demanding (at the time of its creation) memory requirements the algorithm lost in competition with other approaches such as REM. 
Bearing in mind significant improvements in RAM available in contemporary computers we have updated the JW algorithm, tested it for the Ising model and demonstrated its performance for sorption in lattice-gas models (a toy model  and aerogel model). 

The efficiency of the JW algorithm stems from the fact that only a single non-local (multiple-flip) jump is needed to cross the free-energy barriers as opposed to multiple replica exchanges in the case of REM. 
Even though each particular replica within  REM is allowed to cross energy barriers by configuration exchange with a replica at a slightly different temperature, REM overall simply swaps the two replicas, and computer resources continue to be used for exploring the same free-energy minima. 
Therefore, if the minimum is deep enough the algorithm based on REM can spend vast amount of time randomly diffusing in the temperature dimension while exploring the same local free-energy minimum.
In contrast, as temperature decreases the JW algorithm naturally drives the system to escape from the energetically unfavourable local minima and therefore, the system is maintained in the equilibrium throughout the entire duration of the simulation.

The principal differences between the two methods, the JW and REM, suggest the types of problems that can be tackled by each approach.
Since the REM replicas tend to explore the local minima for extended periods of time throughout the simulation, the method is well suited for tasks such as mapping of the entire free-energy landscape, weight factor estimation for multicanonical algorithms and, more generally, for investigating the properties of the whole state space of the system.
The JW algorithm is more appropriate for studying equilibrium behaviour of the system.
As temperature decreases, the system simulated by the JW algorithm naturally approaches the most favourable explored minimum, thus this method is well suited to search for the global free-energy minimum and investigate system's behaviour at the lowest temperatures.
Since significantly less CPU resources is required by the JW algorithm as discussed in Sec.~\ref{sec:method}, it can either enable to study systems that have previously been prohibited by the lack of computing power, or to allow a more efficient investigation of the parameter space, provided that the analysis is mainly focused on the equilibrium behaviour of the system. 
In particular, the JW MC algorithm can be used on a single computing thread to investigate equilibrium properties of a system, especially near discontinuous  phase transitions where the standard single-flip MC simulations are inefficient due to critical slowing down of dynamics and REM is too expensive in terms of computational resources. 

Successful application of the updated JW algorithm to lattice-spin and lattice-gas models (see Sec.~\ref{sec:Results}) might promise its relevance to other models facing quasi-ergodic behaviour, such as spin-facilitated models for glassy dynamics~\cite{Ritort_03:review} and  lattice models for proteins~\cite{Wust_2012}. 
Another possible area for application of the updated JW method could be in studying the systems exhibiting weak ergodicity breaking~\cite{Metzler_2014:review} with non-Boltzmann limiting distributions~\cite{Bel_2005}.  

The description of the improved JW MC method was intentionally given 
as simple and general as possible.  
Therefore, despite certain advantages in its current form, the algorithm  
leaves much room for further optimisation and improvements. 
Moreover, due to key similarities with REM, numerous developments that studies of REM provided throughout the last two decades are readily available to implement for the updated JW algorithm (e.g. the temperature set $\{T_\alpha\}$ and jump frequency optimisation as well as combinations with Wang-Landau, multicanonical or simulated tempering algorithms).

\bibliography{./archive_snt}

\section*{Acknowledgements}

The authors would like to thank Sam Niblett for his help in writing aerogel data structures generating software. The work has been supported by a grant from EPSRC.

\section*{Author contributions statement}

Z.R. and S.T. designed the research. Z.R. performed the research. Z.R. and S.T. prepared the manuscript. 

\section*{Additional information}

The authors declare no competing financial interests.

\end{document}